\documentclass[dvips]{arxstspdf}
\usepackage{graphics}
\usepackage{flushend}

\volume{24}
\issue{4}
\pubyear{2009}
\firstpage{388}
\lastpage{397}
\doi{10.1214/08-STS280}

\begin{document}
\begin{frontmatter}

\title{The Role of Family-Based Designs in Genome-Wide Association Studies}
\runtitle{Family Designs in GWAS}

\begin{aug}
\author[a]{\fnms{Nan M.} \snm{Laird}\ead[label=e1]{laird@hsph.harvard.edu}\corref{}}
\and
\author[a]{\fnms{Christoph} \snm{Lange}\ead[label=e2]{lange@hsph.harvard.edu}}
\runauthor{N. M. Laird and C. Lange}

\affiliation{Nan M. Laird is Professor and Christoph Lange is Associate
Professor}

\address[a]{Nan M. Laird is Professor and Christoph Lange is Associate
Professor,
Department of Biostatistics, 655 Huntington Ave,
Boston, Massachusetts 02115, USA \printead{e1}; \printead*{e2}.}

\end{aug}

% ABSTRACT
\begin{abstract}
Genome-Wide Association Studies (GWAS)
offer an exciting and promising new research avenue for finding
genes for complex diseases.  Traditional case-control and cohort
studies offer many advantages for such designs.  Family-based
association designs have long been attractive for their robustness
properties, but robustness can mean a loss of power.  In this paper
we discuss some of the special features of family designs and their
relevance in the era of GWAS.
\end{abstract}

% KEYWORDS

\end{frontmatter}

%s1 ###
\section{Introduction}

The potential of genome-wide association studies (GWAS) to enable an
unbiased search for disease loci across the entire human genome
provides us with an unprecedented research opportunity in genetics.
Interrogating several hundred thousand SNPs across many subjects at
the same time raises many statistical challenges in the design and
analysis of these studies.  Genotyping on such a scale requires new
methodology for handling data quality issues; likewise, association
tests are computed for hundreds of thousands of markers, whose
results have to be adjusted for multiple comparisons. The magnitude
of these problems raises the question of whether the new technical
ability to genotype such dense SNP sets will translate into the
identification of novel genetic disease loci or whether the
technical advance remains under-utilized.

A popular way to address the multiple testing in genome-wide
association studies has been to design studies with a sample size of
several thousand subjects that are large enough that realistic
effect sizes can be detected, assuming that the test results will be
corrected for multiple testing using the Bonferroni approach.
However, such large studies come at a price. By putting together
samples of several thousand subjects, phenotypic and genetic
heterogeneity will be encountered in the sample. Further, since the
need for large sample sizes also influences the study-design choice,
the most commonly used design choice is a case-control sample of
unrelated individuals with minimal or no covariates.  Another
popular approach is a population-based design of unrelated
individuals without ascertainment condition related to the outcome
of interest (e.g., studying obesity in a general population sample).
In any event, the ascertainment of subjects and collection of their
phenotypic data is rarely carried out specifically for the GWAS;
rather, the expense of the genotyping has led investigators to rely
on samples previously collected and phenotyped for other studies, in
some cases, large family samples that have been previously collected
for other genetic studies.   Although the cost of genotyping is
dropping rapidly, the cost of genotyping still tends to drive study
design and make power considerations very crucial in the design.

An alternative approach to population-based or case-control studies
of unrelated individuals is family-based studies.  Family-based
studies were used in association studies originally to provide
protection against spurious association arising with population
substructure. Family designs offer some unique advantages at the
design and analysis phase of a GWAS.

Their complete robustness against heterogeneity at a phenotypic and
genetic level allows the joint analysis of arbitrarily large and
diverse samples with family designs, an advantage in the GWAS
setting.   As we will discuss in Section \ref{sec3}, they have both
drawbacks and benefits over conventional designs when genotyping
errors are present.  We will also discuss two-stage test strategies
for family designs that maintain the original robustness of the
approach, while achieving power-levels that are similar to those of
population-based studies.

Our objective in this paper is to first describe some of the special
features of family-based designs that make them attractive for
association studies, then focus particularly on their use in GWAS's
with regard to genotyping errors and potential for addressing the
multiple comparison problem.

%f1 ###
\begin{figure}[b]

\includegraphics{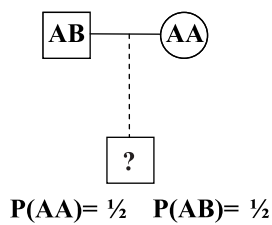}

  \caption{Trio design.}\label{fig1}
\end{figure}

%s2 ###
\section{Overview of Family Designs for a Single Marker}\label{sec2}

It has long been recognized that various sorts of population
substructure can distort tests of association because different
populations may have different disease rates, and/or genotype
frequencies\break (\cite{Devlin1999}; Pritchard, Stephens  and Donnelly (\citeyear{Pritchard2000});
\cite{Whittemore2006}).  Family designs for genetic association
studies were originally suggested (\cite{Falk1987}; \cite{Ott1989};
\cite{Spielman1993}) as a way of avoiding spurious association due
to population substructure. The classic paper by Speilman, McGinnis
and Ewens (\citeyear{Spielman1993}) on the Transmission
Disequilibrium Test (TDT) has contributed much to their general
popularity.  There are many variations on the family design, but the
simplest and generally most powerful design consists of selecting
affected offspring and their parents, and genotyping the trio.
Essentially, having the genotypes of the parents enables one to take
advantage of ``Mendelian Randomization'' to avoid the need for an
explicit control group. Under the null hypothesis of no association
between the disease and the marker, each parent transmits one of
their two alleles to each offspring, at random with probability
50/50  and independently of the other parent and of any other
offspring.  For the example in Figure \ref{fig1}, the mother can only
transmit the A allele, but the father can transmit either A or B
with probability 50/50. This holds whenever there is no selection of
the offspring related to the marker in question.  Thus, when the
parent's genotypes are known, one can easily calculate the
distribution of the offspring genotypes under~$H0$.  This
distribution is used to construct tests of the null hypothesis.  The
observed and expected counts can be used to construct an asymptotic~$\chi^2$
test (\cite{Ott1989}; \cite{Spielman1993}) or exact tests
can be used (\cite{Lazzeroni1998}). Because parents transmit
independently to different offspring, multiple affected siblings can
be used, resulting in a potential savings in genotyping costs.  With
more common diseases, using transmissions to unaffected siblings may
also be beneficial\break (\cite{LangeLaird2002}).

\subsection*{A Class of Score Tests for Family Designs}
 A more precise
statistical argument regarding the robustness of the family designs
can be made by considering the basis for the TDT test.  The simple
TDT test is a score test, based on the likelihood of the offspring
genotypes, conditioned on the offspring trait and the parental
genotypes (\cite{Schaid1996}).  To develop this likelihood in a
general setting, let $P$ denote the parental genotypes of a trio,
$Y$ denote the trait of the offspring (here the trait can be
arbitrary), and let $X$ denote some numerical coding for the
offspring genotype, for example, number of A alleles or a dummy variable
coding for a recessive or dominant genetic model.  Further, let
$f(Y|X, P, \theta)$ denote the probability density of the offspring
trait, conditioned on the offspring genotype, the parental genotype
and a vector of unknown parameters, $\theta$.    In genetic
terminology, $f(Y|X, P, \theta)$ is the penetrance function and
specifies the genetic disease model. Generally, $f(Y|X, P, \theta)$
is assumed not to depend directly on the parental genotypes when
offspring genotypes are in  the model, but we leave them in for
generality. The vector $\theta$ will contain both association
parameters, say, $\beta$, and nuisance parameters, say, $\alpha$,
which will describe other aspects of the trait distribution. In
particular, we parameterize so that $f(Y|X, P, \theta) = f(Y|X
\beta, P, \alpha)$, and under the null, $\beta = 0$, so that $f(Y|X,
P, \beta=0, \alpha) = f(Y|P, \alpha)$, that is, the distribution of the
trait does not depend on the marker genotypes of the offspring under
the null. Further, let $f(X|P)$ be the probability density of the
offspring genotype conditioned on parental genotype. Note that the
latter is completely known and determined by Mendel's laws, whereas
the former reflects our alternative hypothesis, and is generally
\mbox{unknown.}\looseness=1

The conditional likelihood for the offspring genotype $(X)$ given
parental genotypes $(P)$ and the offspring trait $(Y)$ is given by
\begin{eqnarray}\label{eq1}
f(X|Y, P, \theta) &= & f(Y|X, P, \theta) f(X|P)\nonumber
\\[-8pt]\\[-8pt]
&&{}\big/\sum f(Y|X, P, \theta) f(X|P),\nonumber
\end{eqnarray}
where summation is over all $X$ compatible with $P$.  An important
feature of conditioning on $P$ is that any nuisance parameters in
the distribution of the parental genotypes, such as allele
frequencies and random mating assumptions, are not needed.  As noted
above, the penetrance function does not depend on $X$ under the
null, and hence cancels out of the likelihood. Thus, the distribution
of $X$ under the null is given simply by $f(X|P)$, which is
completely determined by Mendel's laws; no assumptions need be made
about the distribution of parental genotypes or about the
phenotypes.   Thus, a score test will have the correctly specified
null distribution as long as Mendel's laws hold, and will be
completely robust to not only population substructure, but to
potential misspecification of the trait distribution as well.

In the TDT, we condition on $Y=1$ and let $X$ denote the number of a
particular allele that an individual has. The model
$p(Y=1|X,P,\theta)$ can take any form, logistic, log-linear, linear,
etc., with $\alpha$ modeling the probability for $X=0$. A simple
form for the penetrance function, which provides a generalization of
the TDT for any phenotype, can be obtained by assuming an
exponential family model for the trait distribution with a
generalized linear model for the mean response
(\cite{Lunetta2000}; \cite{Liu2002}; \cite{Dudbridge2008}).  In
this case, the score takes the special form of a type of covariance
between the trait and the marker:
\begin{equation}\label{eq2}
U = \sum \bigl[ \bigl(Y - E(Y) \bigr) \bigr] [X - E (X|P ) ],
\end{equation}
where summation is over all trios.  Here $E(Y)$ is the mean trait
under $H0$ and may depend upon the unknown nuisance parameters
$\alpha$, and $E(X|P)$ is computed using only Mendel's laws.  An
asymptotic $Z$ (or $\chi^2$) test statistic is formed by normalizing
(\ref{eq2}) by the square root of  $\sum (Y - E(Y) )^2 \operatorname{var}(X|P)$,
where $\operatorname{var}(X|P)$ can also be computed simply from Mendel's Laws.
Alternately, exact tests using\break Mendel's laws to compute $f(X|P)$ can
be easily\break calculated (\cite{Lazzeroni1998} and\break
\cite{Schneiter2005}).

A potential barrier to constructing score tests in this general case
is in estimating the nuisance parameters $\alpha$. Standard
likelihood ratio methods cannot be used here, because under the
null, the likelihood does not depend on $\theta$ and the $\alpha$
parameters cannot be estimated.  The case of trios, where all
offspring are affected $(Y =1)$, is special in this regard.  Here, $Y
- E(Y)$ is constant for everyone, and  because we condition on $Y$,
the score test can be reformulated as
\begin{equation}\label{eq3}
 U = \sum [X - E (X|P ) ].
 \end{equation}
It is easily seen that this score test yields the TDT when $X$ is
coded to count the number of alleles of interest
(\cite{Schaid1996}).

If we include unaffected offspring $(Y=0)$ as well as affected, then
equation (\ref{eq2}) still holds, but the test now depends upon estimating
the prevalence $E(Y)$ because $(Y-E(Y))$ is not constant.  If
selection of subjects depends upon disease status, then prevalence
cannot be estimated from the sample data, but often some a priori
information is available.   In the more general case of measured
phenotypes, the test depends on the specified disease model via the
nuisance parameters implicit in $E(Y)$ and remains valid regardless
of choice of disease model provided Mendel's laws hold.  While model
choice can affect power (\cite{LangeLaird2002};
Lange, DeMeo and Laird (\citeyear{LangeDeMeo2002})), choice of the wrong disease model does not
affect robustness, as the test is conditioned on the trait. When
samples are selected on the basis of the disease trait, as is
generally the case with dichotomous traits, the nuisance parameters
cannot be estimated from the data; methods for specifying $E(Y)$ have
been suggested (\cite{Lunetta2000}; \cite{LangeLaird2002};
\cite{Lu2007};\break \cite{Dudbridge2008}).

\subsection*{Missing Parental Information}
Missing parental genotype
information is a common problem, especially for later onset
diseases.\break  There have been several approaches suggested for handling
missing parents, including estimating a\break model for the parental
genotypes distribution, and using joint likelihood ratio tests
(\cite{Weinberg1999}) or using score tests which average over the
estimated distribution of the parental \ genotypes \
(\cite{Clayton1999}) for families with missing parents.  These
approaches are not guaranteed to retain robustness to population
substructure, especially since both approaches generally make
simplifying assumptions concerning the distribution of the parental
genotypes (e.g., common allele frequencies and Hardy--Weinberg
equilibrium); see \citet{Dudbridge2008}. Alternatively, when
siblings are sampled, $f(X|P)$ can be replaced in the above
equations by $f(X|S)$, where $S$ denotes the sufficient statistic
for parental genotype\break (\cite{Rabinowitz2000}).  Being the
sufficient statistic, $f(X|S)$ again does not depend upon\break a model
for parents' genotype distribution, and the score test remains fully
robust.  The distributions are simple to enumerate, and tests based
on (\ref{eq1})--(\ref{eq2}) with $f(X|P)$ replaced by $f(X|S)$ if parents are
not\break
available can be implemented in the FBAT\break
\href{http://www.biostat.harvard.edu/\textasciitilde fbat/}{www.biostat.harvard.edu/\textasciitilde fbat/}
or PBAT
\href{http://www.biostat.harvard.edu/\textasciitilde
clange}{www.}\break
\href{http://www.biostat.harvard.edu/\textasciitilde clange}{biostat.harvard.edu/\textasciitilde clange}
software packages. However,
the power of these tests can be much reduced, depending upon the
number of additional siblings available.  We refer to the FBAT test
to describe this general class of score tests which extends the TDT
to other traits and other family designs.

In summary, conditioning on both the parental genotypes and the
offspring traits ensures robustness against misspecification of the
disease model, and to the distribution of offspring genotypes under
the null.   The general approach has been extended to handle
multiple siblings (\cite{LangeLaird2002} and
\cite{LangeDeMeo2002}), missing parents (\cite{Rabinowitz2000}),
multiple traits (\cite{Lange2003}),\ \  haplotypes \ \
(\cite{Horvath2004}) and multiple markers (\cite{Xu2006};\break
\cite{Rakovski2007}).

\subsection*{Comparative Power Issues: Single Marker Case} By and large,
most approaches for analyzing\break GWAS studies, conventional or family
designs, begin by testing each marker separately, and then do an
adjustment for multiple comparisons to determine genome-wide
significance and/or select promising\break SNPs or regions for further
study based on rankings of some sort.  There have been several
proposals for alternative methods of testing to increase power in
the face of multiple testing, as we will discuss in Section \ref{sec5},
but,
by-and-large, the genome-wide power of a GWAS is usually estimated
by calculating power for a single marker, using some appropriate
alpha-level to adjust for multiple comparisons; thus, comparative
power issues for single markers translate directly to power
calculations for genome-wide studies.

We note that this one-marker, one-test approach is in strong
contrast to genome-wide linkage scans, where one can at least
approximate the null distribution of the test statistic across the
genome, for example, maximized lod-score, under the null hypothesis of no
linkage (\cite{Feingold1993}).  With dense association scans, the
unknown pattern of LD precludes specification of the joint
distribution of the association test statistics under the null of no
association. In principle, using permutation tests in case-control
studies can considerably improve the probability of at least one
positive finding, but the magnitude of the computations are
prohibitive in a GWAS with hundreds of thousands of SNPs.  An
exception to the one test per typed SNP are methods which
incorporate information from the Hapmap to impute non-typed SNPs,
gaining additional power via testing a denser marker set
(\cite{Marchini2007}). Thus far, this approach has been limited to
case-control data and investigation of methodology for family
designs is desirable.

\begin{figure}[t]

\includegraphics{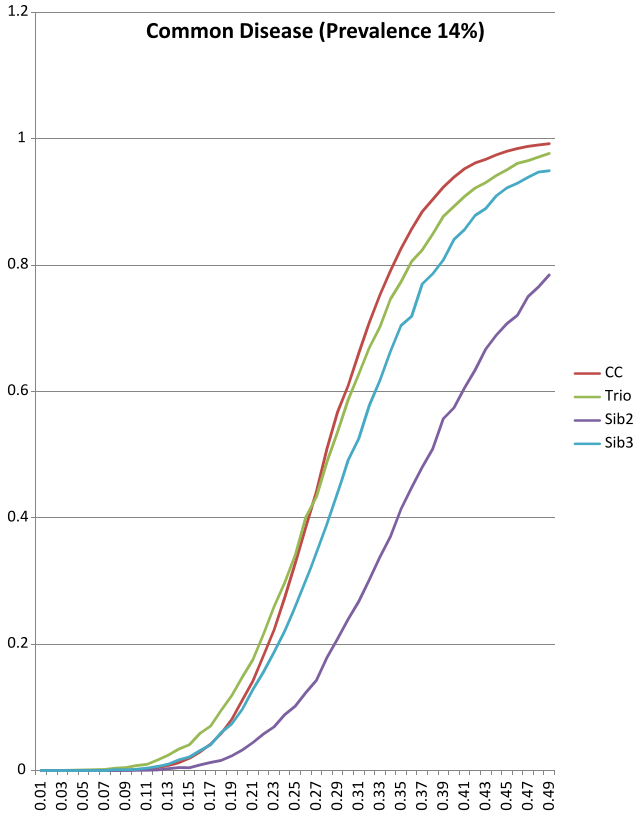}

  \caption{Power for a common disease: 14\%.}\label{fig2}
\end{figure}

\begin{figure}[t]

\includegraphics{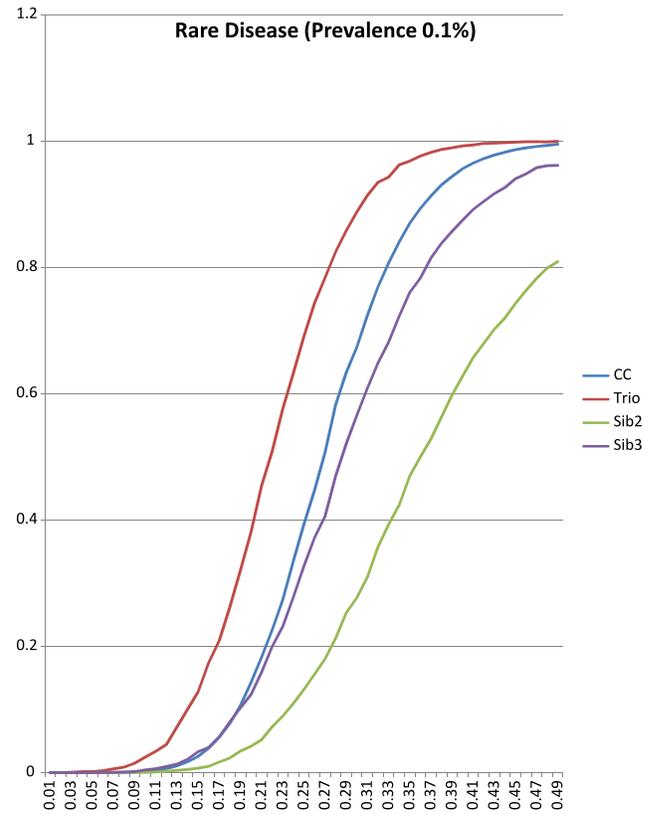}

  \caption{Rare disease: 1\%.}\label{fig3}
\end{figure}

Family-based tests, being conditional tests, are robust and
essentially model free, but the price of such robustness is some
cost in terms of power.  There are some cases, and some designs,
however, where the power is essentially equivalent, as was shown for
rare disease and the additive model in Laird and Lange
(\citeyear{Laird2006}). Here we consider power comparisons for the
recessive model with an $\alpha$-level of 0.00001 to more nearly
reflect a GWAS testing situation.  Figures \ref{fig2} and \ref{fig3} compare the power
of four different designs: case-control, trios, discordant sib pairs
(DSP) and discordant sib trios (DST; at least one discordant sib
pair and one other sibling), for a rare disease and a common one.
The odds ratio is 1.75 in both cases, and the number of affected
(1500) is the same for each design, although number of genotypes
required can be different depending on design. The DSP design is
always very inefficient, whereas DST can do well with more common
disorders.  For the recessive model, the power of the case-control
design
and the trio design are virtually identical
for common diseases (e.g., prevalence 14\%), with minor advantages for trio designs for low allele
frequency and minor advantages for the case-control design for common alleles. However, for rare
diseases, the trio design is much for powerful than the case/control design. The reason for the
relative power loss of the case/control design is that, for rare diseases, the differences between the
genotype distribution of healthy controls and the genotype distribution of the general population
are minimal and the contribution of the controls to the power of the test statistic diminishes. For
the trio design, we use only cases and, consequently, such designs do not suffer this relative power
loss for small prevalences.
%is very low until allele frequency approaches 0.4, regardless
%of prevalence. The family designs are strikingly more powerful in
%the allele frequency range 0.1 to 0.25. The reason for this is that
%both cases and  controls in a case control study have low allele
%frequencies. For the family design, we use only cases and only
%informative families; under the alternative hypothesis, both the
%expected number of informative families and the non-centrality
%parameter are substantial when we condition on affection status and
%parental genotypes.
The power results for the trios differ slightly
by prevalence because we base our model on the odds ratio rather
than the relative risk model.  We provide some simple algebraic
calculations in the \hyperref[app]{Appendix} to illustrate this point.

%f2 ###

%f3 ###

%s3 ###
\section{Quality Control/Data Cleaning in Family
Designs}\label{sec3}

The large amount of genotyping required for a GWAS is accomplished
via specially designed genotyping platforms commonly called
SNP-chips.   Genotyping errors include several types of failures
that can occur in the genotyping process; these can result in either
missingness and/or misclassification of genotypes.  The raw data of
a single genotype for a single individual is a pair of measured
intensities for each allele; the intensities are translated into
genotypes, generally using some type of statistical clustering
algorithm, referred to as the `genotype calling algorithm.'  Perhaps
due to poor DNA quality or design issues of the SNP-chip, the sample
may simply fail to provide intensities or the intensities do not
separate into the three possible genotype clusters, making it
impossible to obtain called genotypes.  These errors all give rise
to missing genotypes. Further missingness arises in the data
cleaning process which is described below.  Misclassification occurs
if the calling algorithm makes a genotype call which is not correct;
the probability for misclassifying a genotype generally increases
with lower minor allele frequencies, and can depend upon the true,
unobserved, genotype.

In the data cleaning step of a GWAS, basic statistical analysis
tools are used as quality control filters to identify SNPs and
probands for which the SNP-chip is not able to provide sufficient
genotyping quality (\cite{Manolio2007}). Such analysis
techniques/filters include tests for departures from the Hardy--Weinberg
Equilibrium, removal of SNPs with low frequencies or low ``call
rates,'' or deletion of individuals with low call rates.  Since the
inclusion of SNPs and probands with misclassified genotypes can lead
to a substantial reduction in power, the data cleaning/filtering
step is one of the most important parts in the analysis of a GWAS.
While there has been much progress in improving genotype calling and
data cleaning algorithms, we can expect that there will continue to
be some level of missing and misclassification in all GWAS's.

When family data are used, an additional quality-control filter that
is applied in the data cleaning step is the removal of Mendelian
inconsistencies.  Mendelian inconsistencies are genotype
configurations in families that violate Mendel's Law. For example,
if a ``B''-allele is observed in a subject whose parents do not
carry any ``B''-alleles, this is an obvious violation of Mendel's
law. Such genotype configurations are excluded from the analysis.
Furthermore, if Mendelian inconsistencies are more frequent for
certain markers or families, this suggests that there are
fundamental problems with the genotyping for these markers/families
and, it is common practice to exclude them from the analysis
altogether. Markers and/or families with more than five Mendelian
inconsistencies are generally removed from the analysis.

For population-based designs, the presence of genotyping errors
resulting in either misclassification and or missing genotypes does
not cause bias under the null provided errors/missingness is
non-differential in cases and controls.  By non-differential, we
mean errors occur irrespective of case or control status.
Genotyping cases and controls separately can lead to differential
genotyping errors, and considerable bias in association tests.  With
non-differential genotyping errors, there is no bias under the null
and the effect of the genotyping error is to simply decrease the overall
power of the GWAS for the population design.

For family-based designs, the effects of genotyping errors are
different. It is a well described phenomenon in the literature
(\cite{Gordon2001},\break \citeyear{Gordon2002};
\cite{Douglas2002};
Sobel,\break Papp and Lange (\citeyear{Sobel2002}); \cite{Kang2004}) that genotyping errors can
cause biased tests with inflated significance levels. With families,
it will be possible to identify some of the misclassified genotypes
by verifying that the offspring's genotype is not plausible based on
the parental genotypes (or in some cases, sibling genotypes).
However, by removing families with transmission inconsistencies from
the association analysis, only a fraction of the genotyping error is
eliminated from the analysis.  In the computation of the test
statistic, this causes a seeming over-transmission of the major
allele, which leads to the anti-conservativeness of the family-based
association test.

Thus, while population-based studies have reduced power in the
presence of genotyping errors, family-based studies will, in
addition to that, have inflated pre-specified  significance levels.
To judge the relative importance of this fundamental difference
between the two study-design types, it is important to consider the
main purpose of GWA studies.  Their goal is the discovery of new
genetic disease loci and their confirmation/replication in
independent studies/samples.  This is typically achieved by
selecting the markers with the smallest $p$-values from the GWA and
trying to confirm/replicate them in independent studies. It is
obvious that the presence of genotyping errors will reduce the
overall power of both design types, either because of reduced power
(case-control) or by both reduced power and inflated type-1 error
(family designs). However, it is unclear for which design type these
effects are more deleterious and careful simulation studies are much
needed to address this issue.

In practice, it will be important to estimate the undetected
genotyping error rate in the data in order to assess the reduction
in overall-power of the GWA study that is attributable to this error
source. Otherwise, if a GWA is unable to identify new loci, it is
unclear whether this is due the actual absence of genetic risk loci
or due to the reduction in overall power caused by poor genotyping
quality. Family-based studies offer a unique possibility to estimate
the undetected genotyping error rate. By looking at the transmission
pattern of the common allele for all genotyped markers in a GWA
study, an overall/\break genome-wide FBAT statistic can be computed and the
undetected genotyping error rate in the study can be estimated
through simulations under various error models (\cite{Fardo2008}).

%s4 ###
\section{Testing Strategies for the Multiple Comparison Problem in Genome-Wide Association Studies}
\label{sec4}

With mapping arrays for more than one million SNPs now available
(\cite{Matsuzaki2004};\break \cite{Di2005}; \cite{Gunderson2006};
\cite{Wadma2006}), genome-wide association studies carry the
promise to identify replicable associations between important
genetic risk factors and most complex diseases. One of the major
hurdles that needs to be addressed in order to make genome-wide
association studies successful is the multiple comparison problem.
Hundreds of thousands of SNPs are genotyped and examined for
potential associations with multiple phenotypes, possibly using
different model assumptions, resulting in potentially millions of
statistical tests.

Initial efforts to resolve this problem with case-control designs
were directed toward multi-stage designs involving multiple
independent samples.  At stage~1, all SNPs are tested in a
relatively small sample and the most significant ones retained for
testing with a larger, independent sample; the winnowing process can be
repeated multiple times.  However, Skol et al.
(\citeyear{Skol2006}) showed that such designs are inherently less
powerful than designs which use all samples for the final analysis
of selected SNPs, even though Bonferroni adjustment must be made for
testing all SNPs. Thus, the desired strategy now for population based
designs is to select a large enough sample (3--5000 cases and an
equal number of controls) to achieve sufficient power for all SNPs
simultaneously, but also utilize independent ``replication'' samples
which are different from the original sample in some distinct way,
for example, non-overlapping populations.

Other strategies to ameliorate the multiple comparisons problem
utilize some ``outside'' information, for example, information from linkage
studies, functional SNPs, etc.  Such approaches include Bayesian
approaches which use prior distributions to specify effects for
markers (\cite{Wakefield2008}), weighted Bonferroni methods which
assign different significance levels to each SNP according to their
``importance or relevance'' (Roeder,  Devlin and\break Wasserman (\citeyear{Roeder2007});
\cite{Eskin2008})
and split-sample approaches (\cite{Wasserman2006};\break
\cite{Song2007}). For family-based association\break tests, the idea of
using ``outside information'' naturally translates to the use of the
information about the association at a population-based level that
is not utilized in the family-based association test.

A general approach to two-stage testing for family designs builds on
the two information sources about association that are present in
family-based designs.  Using the notation introduced in Section
\ref{sec2}
for the distribution of $X$ and $Y$, the joint distribution for $X$,
$Y$ and $P$ (or, equivalently, $S$)  can be partitioned into two
statistically independent components (\cite{Laird2006}),
\begin{equation}
\quad f(X,Y,P|\Phi,\theta) = f(X|Y,P,\theta) f(Y,P|\Phi,\theta),
\end{equation}
where $\Phi$ represents additional parameters required to model the
parental genotype distribution, for example, genotype frequencies and
possible non-random mating.  Note that both components,
$f(X|Y,P,\theta)$ and $f(Y,P|\Phi,\theta)$ will have information
about $\theta$, but the information from $f(Y,P|\Phi,\theta)$, will
depend on the parental genotype distribution, and can be sensitive to
population substructure.

For the first step of the testing strategy, the screening step, we
use the information in $f(Y,P|\Phi,\theta)$, to estimate the
association parameters; the second, or testing step, uses
$f(X|Y,P,\Phi,\theta)$. The likelihood decomposition implies that
both steps of the testing strategy are independent. The ``evidence
for association'' estimated from $f(Y,P|\Phi,\theta)$ can be
utilized in the testing stage, without having to adjust the test for
the estimation of the genetic effect size in the first stage.
Several methods have been suggested to exploit this relationship in
developing testing strategies which use both forms of information in
order to increase power, while retaining robustness of the test.

Van Steen et al. (\citeyear{VanSteen2005a}) originally proposed a
version of this two-step testing strategy for the analysis of
quantitative traits. First, an effect size is estimated for each SNP
by regressing the offspring phenotype $Y$ on $E(X|P)$;  this effect
size is used to calculate the estimated power of the FBAT statistic
for each SNP (\cite{LangeLaird2002}). Some number of top ranking
SNPs (10 or 20) were selected for testing with the FBAT statistic at
the second stage. Because of the independence, both steps can be
applied to the same data set without having to adjust the overall
significance level for the multiple usage of the data. An extension
by Ionita-Laza et al. (\citeyear{Ionita2007}) proposed testing all SNPs
at the second stage using weighted Bonferroni.  Extensions of this
testing strategy are available for using parental phenotypes and
arbitrary structures at the screening stage (\cite{Feng2007}) and
for case/control designs (\cite{Zheng2007}).

The Van Steen approach has three key advantages: (1) The method
achieves statistical power levels which can be substantially higher
than those of standard family-based approaches and is thereby able
to establish genome-wide significance with\break smaller/more realistic
sample sizes (\cite{VanSteen2005b}; \cite{Ionita2007};
Feng, Zhang and Sha (\citeyear{Feng2007}); \cite{Zheng2007}). (2) The Van Steen algorithm
maintains the separation between the multiple testing problem and
the replication process. Replication attempts in different studies
are reserved for the generalization of the established associations
and the assessment of heterogeneity between study populations.
(3)~Since genome-wide significance is established in the first data set,
the number of SNPs that is pushed forward to true replication in
other populations is generally very small and does not require a
large budget, which makes simultaneous replication attempts in
multiple samples feasible. Extensive simulation studies have shown
that 2-stage testing strategies that utilize both sources of
information about the association can help family-based studies to
achieve power levels that are similar to those of population-based
studies, while maintaining the original advantages of family-based
study, that is, complete robustness against confounding.

By looking at the distribution of parental mating types in
ascertained samples $f(P|Y, \Phi, \theta)$, Murphy et al.
(\citeyear{Murphy2008}) extended the general approach to the
trio-designs in which all probands are affected $(Y=1)$.  Even here,
the application of 2-stage Van Steen-testing strategies can lead to
meaningful power improvements over the standard TDT.  Other
possibilities for utilizing the information from the screening step
include specifying ``tuning-parameters'' in the FBAT-statistic
(\cite{LangeStatApp2004}; \cite{Jiang2006}) so that the power of
the FBAT test is maximized.

%s5 ###
\section{Discussion}\label{sec5}

Family designs have historically been popular because of their
robustness to population substructure.   An additional, often
unappreciated, feature of family-designs which is important with
measured or time-to-onset outcomes is their robustness to model
specification, and the ability to utilize the population
\mbox{information}
to specific unknown parameters in the model.  With the availability
of modern SNP chips, and genotyping of thousands of subjects on
hundreds of thousands of markers, we now have the potential to
identify the genetic backgrounds of individuals, and utilize that
information to control for confounding by population substructure in
case-control studies (\cite{Roeder2008}).  An important question is
whether or not there is a need for family designs in the era of
GWAS, given the potential to resolve difficulties with population
substructure in case control designs.   Additional studies and
experience with actual studies are needed to compare the performance
of family designs and adjusted case-control designs in GWAS
settings.

Hampered by limitations in terms of power in many scenarios, and by
the difficulty of recruitment, family-based designs certainly cannot
be considered as the gold standard approach in genome-wide
association studies.  However, given the unique properties and
features of a family design, they will continue to play a pivotal
role in large scale association studies.

In multi-stage genome-wide association studies, family-based studies
should be utilized as one of the stages as early as the budget
permits its implementation. Their complete robustness against both
genetic confounding and misspecification of the phenotypic model
provides them with an important role in the process of replicating
and validating findings of the discovery step.  Given the
unavoidable genetic and phenotypic heterogeneity in large-scale
multi-stage genome-wide association studies, this feature of
family-based association tests is crucial and should not be ignored.
If the budget permits the additional genotyping cost, family-studies
can be a favorable choice for the first stage of a genome-wide
association study. There, family-based studies can be designed so
that they have equivalent power to\break population-based studies and, at
the same time, offer a unique combination of additional analysis
features and robustness properties.

While the analysis features of family-based designs make them an
attractive choice in the design phase of genome-wide studies, their
abilities to assess the magnitude of the hidden genotyping error
should always be utilized, even with case/control designs. By
genotyping a small number of families on the same platform with the
case/control samples, researchers can examine the genotyping quality
of the data after the QC process and assess the true power of the
study.

\appendix
\section*{Appendix}\label{app}

Here we do some simple calculations which illustrate the power
differences between case-control and trio designs.  The basic idea
is to calculate the expected value of the corresponding $Z$
statistics under the alternative.  To make the calculations simple,
we use a relative risk model, and we assume that allele frequency,
the relative risk and prevalence are small. We use the following
notation: $p$ $=$ disease allele frequency, $\rho$ $=$ relative risk,
$K$ $=$ prevalence, $r = P(Y=1|X=0)$, where $Y = 1$ indicates disease,
and $X = 1$ indicates the recessive genotype.  Assuming the Hardy--Weinberg
Equilibrium holds in the population, $P(X=1) = p^2$ and $K =
\rho r p^2 + r(1-p^2) \Rightarrow r = K /  ( \rho p^2 +  (
1-p^2  )  )$.

\renewcommand{\theequation}{\arabic{equation}}
\setcounter{equation}{4}

For the case-control design, we compute
\begin{eqnarray}
p_{\mathrm{cases}} &=& P(X=1|Y=1)\nonumber
\\
&=& r\rho p^2 /K\quad \mbox{and}\nonumber
\\[-8pt]\\[-8pt]
p_{\mathrm{controls}} &=& P(X=1|Y=0)\nonumber
\\
&=&
(1-r\rho) p^2 /(1-K),\nonumber
\end{eqnarray}
and letting $\bar{p} =
(p_{\mathrm{cases}} - p_{\mathrm{controls}})/2$, we have that the expected $Z$ is approximately
\begin{equation}
E(Z) = \surd N(p_{\mathrm{cases}} -
p_{\mathrm{controls}})/ \sqrt{2\bar{p}  ( 1-\bar{p}
 )},
 \end{equation}
 where $N$ is the number in each group. For
$N=1500$, $p = 0.1$, $K=0.01$ and $\rho = 1.75$, this gives
$p_{\mathrm{cases}} \approx 0.0174$,
$p_{\mathrm{controls}} \approx 0.0099$ and $E(Z) \approx
1.75$, which corresponds to the notion of zero power if $\alpha =
0.00001$.

For the trio design, we consider the 2 informative mating types,
that is, 2 heterozygous parents (Type 1) and one heterozygous parent
and one rare homozygous parent (Type 2). Under the alternative
hypothesis, the expected number of families for each mating type can
be calculated by
\begin{longlist}
\item[Type 1:] $r p^2 (1-p)^2 (\rho + 3) N/K$,

\item[Type 2:]  $2r p^3 (1-p) (\rho + 1) N/K$.
\end{longlist}

Next, we compute the Mendelian residuals which are defined as the
expected marker score under the alternative hypothesis minus the
expected marker score under the null-hypothesis for both mating
types:
\begin{longlist}
\item[Type 1:] $3 (\rho-1) / 4(\rho + 1)$,

\item[Type 2:] $(\rho-1) / 2 (\rho+1)$.
\end{longlist}

The variance of the mating-types used in the denominators of the
FBAT statistics are given by 3$/$16 and 1$/$4 respectively.

Then the expected FBAT-statistic for a recessive model under the
alternative hypothesis is given by
\begin{equation}
\quad E(Z) = \frac{2p (\rho -1) \sqrt{N{(r/K)} (1-p)} (3+\rho)}{\sqrt{\rho (p+3) - 5p +
9}}.
\end{equation}
 For
the parameters given above, this equals $Z=4.56$, which results in
the observed power levels of the plot for $K=0.01$.

\section*{Acknowledgments}
Supported in part by
a grant from the National Institute of Mental Health.

\end{document}